\documentclass[pss]{wiley2sp} 
\usepackage{amsmath}

\begin{document}

\title{Half-metallic ferrimagnet formed by substituting Fe for Mn in semiconductor MnTe}

\titlerunning{Half-metallic ferrimagnet formed by substituting Fe for Mn in semiconductor MnTe}

\author{%
  Li-Fang Zhu\textsuperscript{\textsf{\bfseries 1,2\Ast}},
  Bang-Gui Liu\textsuperscript{\textsf{\bfseries 1,2}}.}

\authorrunning{Li-Fang Zhu and Bang-Gui Liu}

\mail{e-mail
  \textsf{lfzhu@aphy.iphy.ac.cn}, Phone
  +86-10-82649438, Fax +86-10-62553698\\
  e-mail
  \textsf{bgliu@aphy.iphy.ac.cn}, Phone
  +86-10-82649437, Fax +86-10-62553698}

\institute{%
  \textsuperscript{1}\,Institute of Physics, Chinese Academy of Science,
Beijing 100190, China\\
  \textsuperscript{2}\,Beijing National Laboratory for
Condensed Matter Physics, Beijing 100190, China}

\received{XXXX, revised XXXX, accepted XXXX} 
\published{XXXX} 


\abstract{%
 A ternary ferrimagnetic half-metal, constructed through
substituting 25\% Fe for Mn in zincblende semiconductor MnTe, is
predicted in terms of accurate first-principles calculations. It
has a large half-metallic (HM) gap of 0.54eV and its ferrimagnetic
order is very stable against other magnetic fluctuations. The HM
ferrimagnetism is formed because the complete moment compensation
in the antiferromagnetic MnTe is replaced by an uncomplete one in
the Fe-substituted MnTe. This should make a novel approach to new
HM materials. The half-metal could be fabricated because Fe has
good affinity with Mn, and useful for spintronics.
%
%
%
}

%
%

\maketitle   

\section{Introduction}
Half-metallic (HM) ferromagnets have attracted much attention
because they have band gaps at the Fermi energy for one electronic
spin channel and are metallic for the other
channel\cite{Wolf,Pickett}. A lot of HM ferromagnetic (FM)
materials have been found
\cite{Groot1,Groot2,Dong,Watts,Jedema1,Jedema2,Coey}. Accurate
first-principles calculations have revealed HM ferromagnetism in
binary transition metal chalcogenides and pnictides in the
zincblende and wurtzite
structures\cite{sh,Xie1,Xie2,Xie3,lbg1,lbg2}. It is exciting that
a Singapore group, stimulated by the theoretical prediction of
zincblende CrTe (z-CrTe)\cite{lbg1,lbg2}, has fabricated z-CrTe
samples of 100 nm thickness\cite{zbCT}. It has also been reported
that half-metallic ferrimagnets can be formed by introducing Cr
antisites in CrAs or CrSb\cite{Galanakis}. It is still highly
desirable to search for novel semiconductor-compatible half-metals
with high Curie temperature for potential spintronic applications
\cite{Fang}.

Magnetic materials with and based on zincblende structure are very
interesting to spintronic applications. Zincble-nde MnTe (z-MnTe)
is one of a few antiferromagnetic (AF-M) semiconductors. Although
MnTe crystallizes into a NiAs phase, the metastable z-MnTe has
been grown by molecular beam epitaxy (MBE) growth
technique\cite{Durbin} and semibulk (about 1 micrometer thick)
film samples of z-MnTe have been fabricated\cite{mumnte} because
z-MnTe is only 0.02eV per formula unit higher in total energy than
the NiAs-type MnTe. Ternary Cr-doped NiAs-type manganese
tellurides, Mn$_{1-x}$-Cr$_{x}$Te, with $x$ being up to 14\%, have
been fabricated, in which the substitution of Cr for Mn leads to a
change from an AFM semiconductor of MnTe to a FM (or
ferrimagnetic) semiconductor of Mn$_{1-x}$Cr$_{x}$Te\cite{Li}.
Therefore, z-MnTe should be an interesting novel approach to
explore promising magnetic semiconductors and HM compounds.

In this paper, we perform first-principles study on structural,
electronic, and magnetic properties of the 25\%-Fe-doped z-MnTe.
The substitution of Fe for Mn results in a transition from the AFM
semiconductor of z-MnTe to the ferrimagnetic half-metal of
Mn$_{3}$FeTe$_{4}$. We understand the mechanism of the magnetism
and the magnetic transition through investigating the atomic and
electronic structures of Mn$_{3}$FeTe$_{4}$ in comparison with
those of z-MnTe.

The remaining part of this paper is organized as follows. In next
section we present our computational detail. In the third section
we shall present our optimized results of crystal structures and
investigate the stability of the ferrimagnetism against magnetic
fluctuations. In the fourth section we shall present the
electronic structures and discuss the mechanism for the
half-metallic ferrimagnetism. Finally we shall make some
discussions and give our conclusion.

\section{Computational detail}

To perform the calculations, we use the package
WIEN2K\cite{Blaha}, which is based on full-potential linearized
augmented plane wave method within the density-functional theory
(DFT)\cite{Hohenberg}. The Perdew-Bur-ke-Ernzerhof 1996
version\cite{Perdew} of the generalized gradient approximation
(GGA) is used for the exchange-correlation potential. Full
relativistic effects are calculated for core states, and the
scalar relativistic approximation is used for valence states. We
investigate the effect of the spin-orbit coupling, but still
present  the results without spin-orbit coupling in the following
because it does not affect our main conclusions. For different
magnetic structures we use different but appropriate k points in
the first Brillouin zones and make the expansion up to \emph{l}=10
in muffin tins. $R_{\rm mt}$$\times$$K_{\rm max}$ is set to 8.5
for z-MnTe and to 7.0 for Mn$_{3}$FeTe$_{4}$ without affecting our
conclusions. The self-consistent calculations are considered to be
converged when the integrated charge difference per formula unit
between input and output charge density is less than 0.0001.

\section{Optimized crystal structures}

Recent inelastic ne-utron-scattering experiment has revealed that
the stable ma-gnetic structure of z-MnTe is collinear type-III AFM
order of Mn spins in a double conventional unit
cell\cite{Hennion,Wei2}, rather than early type-I AFM order in
single conventional unit cell\cite{Wei1} or noncollinear type-III
AFM order suggested in terms of previous neutron-diffraction
result\cite{mumnte}. Therefore, we consider only collinear spin
configurations in the following.

\begin{figure}[!htbp]
\begin{center}
\includegraphics[width=.45\textwidth]{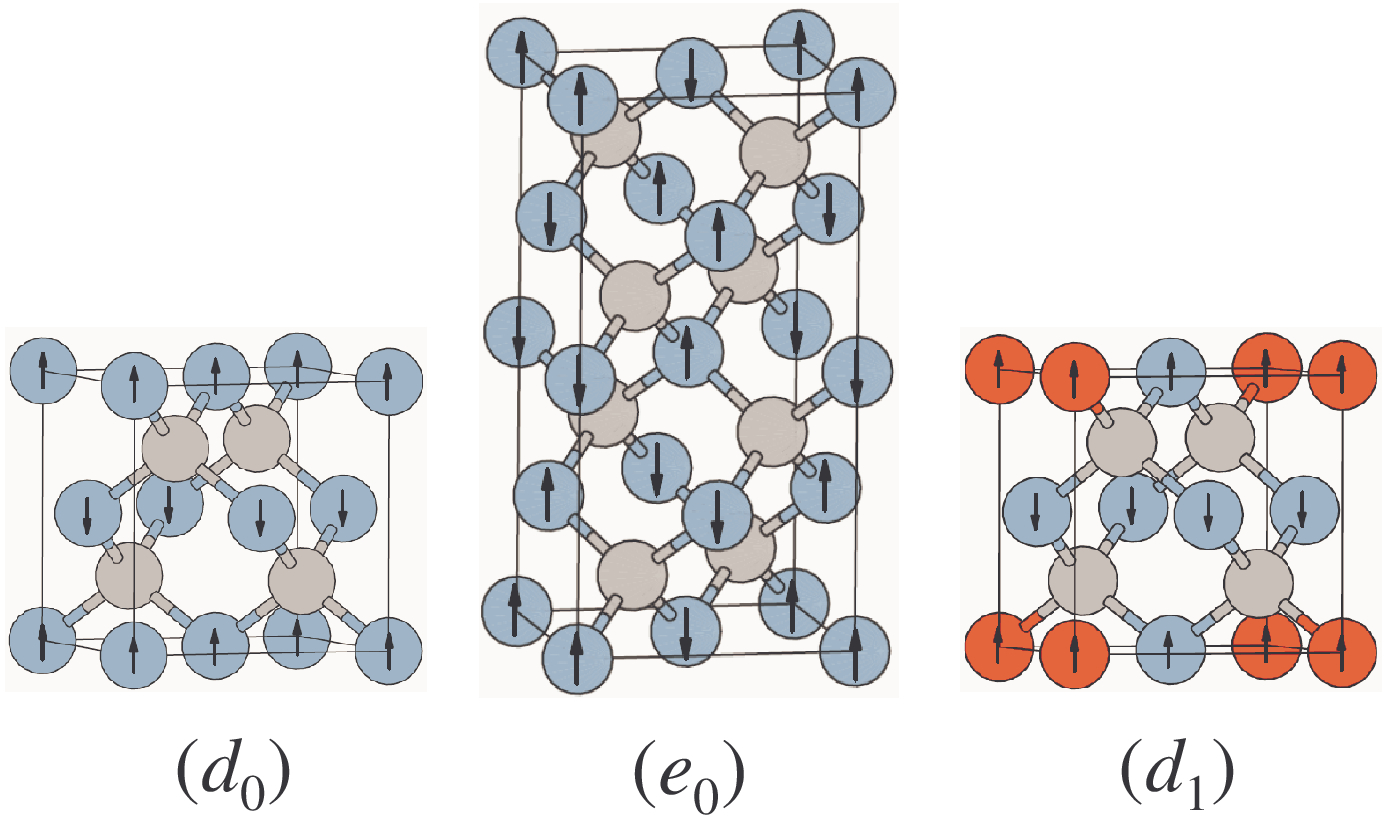}
\caption{(color online). AFM-I ($d_{0}$) and AFM-III ($e_{0}$)
structures of zincblende MnTe and the most stable structure
($d_{1}$) of Mn$_{3}$FeTe$_{4}$. The black (red) ball denotes Fe
(with arrow), the grey (blue) one Mn (with arrow) or Te. The
arrows represent the spins on the sites.} \label{fig1}
\end{center}
\end{figure}

Five spin configurations $a_{0}$, $b_{0}$, $c_{0}$, $d_{0}$ and
$e_{0}$ can be constructed for z-MnTe. $a_{0}$ is a FM structure
with four Mn moments being in parallel. $b_{0}$ and $c_{0}$,
 obtained by reversing one Mn moment respectively at the
face-center and on the vertex of $a_{0}$, are ferrimagnetic
structures and equivalent with each other. $d_{0}$ and $e_{0}$ are
 the type-I AFM (AFM-I) structure with single
conventional cell\cite{Wei1} and the type-III AFM (AFM-III)
structure with double unit cells\cite{Hennion}, respectively, as
shown in Fig. \ref{fig1}. By substituting Fe for Mn on the vertex
in the single MnTe cell of $a_{0}$ $\sim$ $d_{0}$, we get
correspondingly four FM (or ferrimagnetic) structures $a_{1}$,
$b_{1}$, $c_{1}$, and $d_{1}$ of Mn$_{3}$FeTe$_{4}$. $d_{1}$ is
the most stable among them and is shown in Fig. \ref{fig1}.
Generally speaking, to get an AFM structure we construct a
supercell of two unit cells and make the moments in one unit cell
opposite to those in the other. $a_{1}$ $\sim$ $d_{1}$ are all
possible FM (or ferrimagnetic) structures one can construct
without enlarging the magnetic unit cell. We construct all
possible AFM structures based on them, and the results for the
most stable AFM structure $e_1$ (as the representative) are shown
in Table \ref{table1}.

\begin{table}[!btp]
\caption{The space groups (SG), the magnetic orders (MO), the
lattice constants ($a$ or $a$/$c$), the relative energy $E_{r}$
(defined with respect to the lowest structure for the same
formula), the absolute value of total magnetic moment ($M$), and
the Kohn-Sham gaps ($E_{g}$) or the HM gaps ($E_{h}$). $E_{r}$ and
$M$ are normalized in terms of those of $d_{1}$ for comparison. }
\begin{center}
\begin{tabular}{ c c c c c c }\hline
                      \multicolumn{6}{c}{z-MnTe}\\
\hline
                      &$(a_{0})$&$(b_{0})$&$(c_{0})$&$(d_{0})$ &$(e_{0})$     \\\hline
SG                    & 216     & 215     & 215     & 111      & 122          \\
MO                    &    FM   &     FM  &     FM  &     AFM  &   AFM        \\
$a$(/$c$ \AA)         &  6.393  &  6.314  &  6.314  &  6.290   &  6.290/12.580\\
${E_{r}}$(eV)         &  0.712  &  0.196  &  0.196  &  0.020   &  0       \\
\emph{M}($\mu$$_{B}$) &  20.000 & 10.00   & 10.00   &  0.000   &  0.000       \\
${E_{g}}$(eV)         &  $-$    &  0.90   &  0.90   &  1.30    &  1.35        \\
\hline \multicolumn{6}{c}{Mn$_{3}$FeTe$_{4}$}\\
\hline
                      &$(a_{1})$&$(b_{1})$&$(c_{1})$&$(d_{1})$&$(e_{1})$     \\
SG                    & 215     & 215     & 111     & 111     &  35          \\
MO                    &  FM     &   FM    &  FM     &    FM   &  AFM         \\
$a$(/$c$ \AA)         &  6.305  &  6.244  &  6.251  &  6.223  &  8.822/12.476\\
${E_{r}}$(eV)         &  0.764  &  0.114  &  0.222  &  0      &  0.125       \\
\emph{M}($\mu_{B}$)   &  18.226 &  11.000 &  9.000  &  1.000  &  0.000       \\
${E_{h}}$(eV)         &  $-$    &  0.21   &  0.16   &   0.54  &$-$
\\\hline
\end{tabular}
\end{center}
\label{table1}
\end{table}

All the above structures, both FM and AFM, are optimized fully.
The moment and electronic structures are calculated with the
lattice constants of the optimized structures. Our calculated
results are summarized in Table \ref{table1}. It is clear that the
most stable structure tends to have a small equilibrium lattice
constant. As is shown in Table \ref{table1}, the two FM structures
($a_{0}$ and $a_{1}$) and the four ferrimagnetic structures
($b_{0}$, $c_{0}$, $b_{1}$ and $c_{1}$), having large magnetic
moments, are unfavorable in total energy. For z-MnTe, AFM-III
$e_{0}$ and AFM-I $d_{0}$, with the total moments being 0, are
favorable in total energy, and $e_{0}$ is 5meV per formula unit
lower than $d_{0}$, being in agreement with experimental fact that
$e_{0}$ is the ground-state phase of z-MnTe with a semiconducting
gap of about 3.2eV\cite{Durbin}. The most stable structure for
Mn$_{3}$FeTe$_{4}$, however, is not any AFM structure, but the
ferrimagnetic structure $d_{1}$ with an absolute total moment of
1.000$\mu_B$. It is lower by 0.125eV per formula unit in total
energy than the lowest AFM structure $e_{1}$.

\begin{table}[!btp]
\caption{The partial magnetic moments ($\mu$$_{B}$) projected in
the muffin-tin spheres of Mn1, Mn2, Fe, and Te atoms and in the
interstitial region (Inter) and the total moment (Total) in the
most stable structures z-MnTe $e_{0}$ and Mn$_{3}$FeTe$_{4}$
$d_{1}$.}
\begin{center}
\begin{tabular}{ c c c c c c }\hline
              &\multicolumn{1}{c}{z-MnTe $e_{0}$}  &\multicolumn{1}{c}{Mn$_{3}$FeTe$_{4}$ $d_{1}$}&\\\hline
Mn1           & 4.180                               &  4.130    \\
Mn2           &-4.180                               & -4.114     \\
Fe            & $-$                                 &  3.157     \\
Te            & 0.000                               &  0.036     \\
Inter         & 0.000                               & -0.095     \\
Total         & 0.000                               & -1.000     \\
\hline
\end{tabular}
\end{center}
\label{table2}
\end{table}

We summarize the partial magnetic moments ($\mu$$_{B}$) projected
in the muffin-tin spheres of Mn1, Mn2, Fe, and Te atoms and in the
interstitial region in the most stable structures z-MnTe $e_{0}$
and Mn$_{3}$FeTe$_{4}$ $d_{1}$ in Table \ref{table2}. The
corresponding total magnetic moments also are presented for
comparison. It is worth noting that there are two Mn1 atoms and
two Mn2 ones in z-MnTe $e_{0}$, but we have one Mn1 atom, one Fe
atom, and two Mn2 atoms in Mn$_{3}$FeTe$_{4}$ $d_{1}$, as shown in
Fig. \ref{fig1}. It is obvious that the partial substitution of Fe
for Mn leads to the transferring of a little magnetic moments from
the Mn atoms to the Te atoms and the interstitial region.
Mn$_{3}$FeTe$_{4}$ $d_{1}$ has a total moment of -1.000
$\mu$$_{B}$ because Fe has one more $d$ electron, or one $\mu_B$
less magnetic moment, than Mn.

\section{Electronic structures and magnetic mechanism}

\begin{figure}[!htbp]
\begin{center}
\includegraphics[width=.45\textwidth]{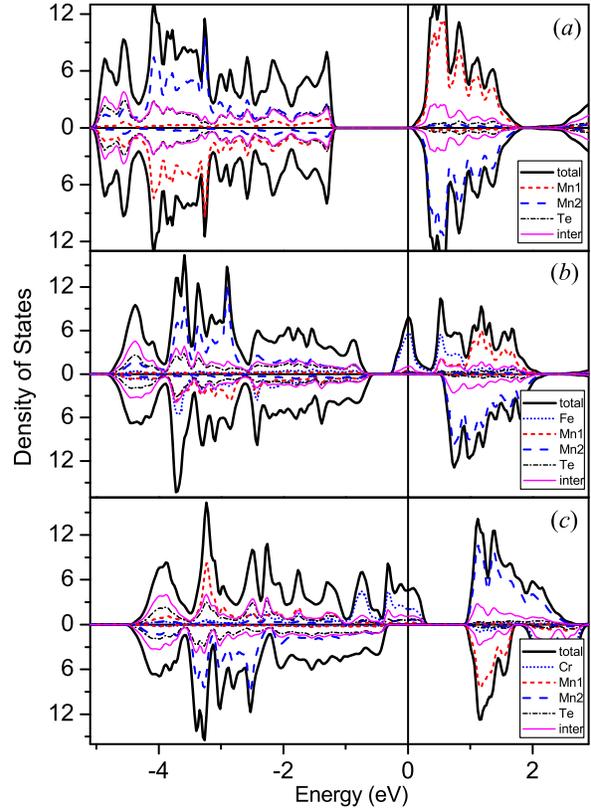}
\caption{(color online). Spin-dependent total (thick solid) and
partial DOS (state/eV per formula unit) for  the AFM-III MnTe
($a$),  the most stable $d_{1}$ structure of Mn$_{3}$FeTe$_{4}$
($b$),  and the most stable structure of Mn$_{3}$CrTe$_{4}$ ($c$).
The upper half of each panel is DOS for majority spin and the
lower one for minority spin. The partial DOS are those in Fe/Cr
(dot), Mn1 (dash), Mn2 (long dash), and Te (dot dash) muffin tins
and in interstitial region (thin solid).} \label{fig2}
\end{center}
\end{figure}

The spin-dependent density of states (DOS) of the AFM-III MnTe are
presented in Fig. \ref{fig2}($a$). The primitive cell of AFM-III
MnTe consists of 2 Mn1 (with spin up), 2 Mn2 (with spin down), and
4 Te atoms. The valence bands are formed by 10 $d$ and 12 $p$
states. The 10 lowest conduction bands originate from Mn $d$
states. The Mn moments are coupled with a superexchange
interaction through the nearest Te atoms, which yields the
antiferromagnetism. The spin exchange splitting is about 4.7eV, as
shown in Fig. \ref{fig2}($a$).

\begin{figure}[!htbp]
\begin{center}
\includegraphics[width=.48\textwidth]{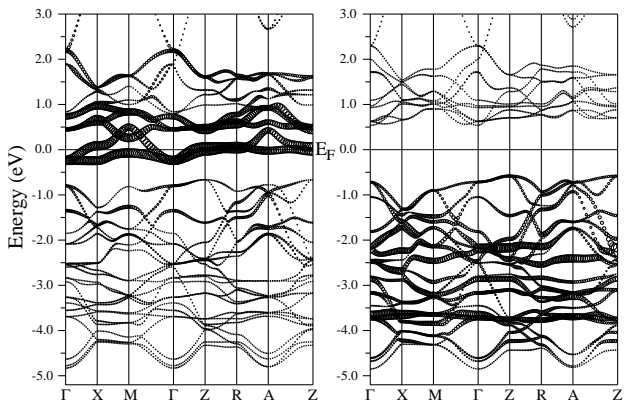}
\caption{Spin-dependent energy bands (plotted with circles) of the
$d_{1}$ structure of Mn$_{3}$FeTe$_{4}$. A larger circle implies
more Fe $d$ character. The left panel is for majority-spin and the
right one for minority-spin.} \label{fig3}
\end{center}
\end{figure}

The spin-dependent density of states (DOS) and energy bands of the
Mn$_{3}$FeTe$_{4}$ are presented in Fig. \ref{fig2}($b$) and Fig.
\ref{fig3}, respectively. The Fermi level $E_F$ is set to zero.
The filled bands between -5eV and -0.6eV, consisting of 10 $d$
states and 12 $p$ states for each spin channel, are similar to
those of the MnTe. The main difference is that there are
partly-filled majority-spin bands across the Fermi level in the
case of the Mn$_{3}$FeTe$_{4}$. The minority-spin bands still have
a gap of 1.12eV, a little smaller than the Kohn-Sham gap, 1.35eV,
of the MnTe. The Mn$_{3}$FeTe$_{4}$ has 0.54eV as its HM gap which
is defined as the smaller of $E^c_{\rm min}$-$E_F$ and
$E_F$-$E^v_{\rm max}$, where $E^c_{\rm min}$ is  the bottom of the
minority-spin conduction bands and $E^v_{\rm max}$ the top of the
minority-spin valence ones\cite{lbg1,lbg2}.

The 25\% Fe substitution for Mn results in a cell consisting of 2
Fe, 6 Mn, and 8 Te atoms. The spin orientations cannot remain the
same as those of AFM-III MnTe because Fe has one more $d$ electron
than Mn. Instead, the magnetic order is reorganized so that the
16-atom cell is divided into two equivalent smaller 8-atom ones
which would have AFM-I structure if we neglect the difference
between Fe and Mn. As a result, we obtain a ferrimagnetic order
because the Fe moment cannot completely compensate the opposite Mn
moment. The substitution does not substantially change the valence
bands, but moves some majority-spin $d$ states downwards with
respect to those of the z-MnTe because the Fe $d$ states are a
little lower than those of Mn $d$ ones in energy. The
majority-spin bands at the Fermi level, belonging to a doublet,
are half-filled because there is only one electron for them. The
HM ferrimagnetism is achieved because we still have a gap across
the Fermi level for minority-spin channel.

We have studied similar 25\%-Cr-doped MnTe, Mn$_{3}$Cr-Te$_{4}$.
Its stable structure also exhibits HM ferrimagnetism. The results
for Mn$_{3}$CrTe$_{4}$  are consistent with Nakamura \emph{et
al}'s through doping 75\% Mn into z-CrTe \cite{Nakamura}. The
spin-dependent density of states for the most stable structure of
Mn$_{3}$CrTe$_{4}$ are also given in Fig. \ref{fig2}($c$). Cr has
four $d$ electrons, one less than Mn. The Cr $d$ states are a
little higher than those of Mn in energy, which results in the
partially occupied Cr impurity bands in the majority-spin bands
and the open gap in the minority-spin bands.

By comparing the DOSs of the Mn$_{3}$FeTe$_{4}$ and
Mn$_{3}$Cr-Te$_{4}$ in Fig. \ref{fig2}, we can explain the origin
of their ferrimagnetism uniformly according to the number of $d$
electrons in the transition metal atoms and the energy levels of
$d$ states. The substitution of Fe for Mn or Cr for Mn changes the
distribution of $d$ states at the fermi level and results in the
ferrimagnetism.

\section{Discussion and conclusion}

All of our presented results are calculated with GGA, although
local density approximation (LDA) yields almost the same results.
It is worth noting that a developed single-ion implantation
technique recently was used to implant dopant ions one-by-one into
a semiconductor\cite{Takahiro}. That is, both the number and the
position of the dopant atoms in the semiconductor are precisely
controlled. As a result, the promising half-metals predicted in
this paper could be realized by using such techniques.

In summary, we have predicted a ternary half-metal
Mn$_{3}$FeTe$_{4}$, constructed by substituting Fe for Mn in
semiconductor z-MnTe, in terms of our accurate first-principles
calculations. The substitution results in a transition from the
AFM semiconductor MnTe to the HM ferrimagnet of the
Mn$_{3}$FeTe$_{4}$. The HM ferrimagnetism is stable against
antiferromagnetic fluctuations. The large HM gap implies a
possible high Curie temperature\cite{Kubler}. The
Mn$_{3}$FeTe$_{4}$ could be fabricated experimentally soon because
of the good affinity of Fe to Mn, and it could be used in
spintronics.

\begin{acknowledgement}
This work is supported  by Nature Science Foundation of China
(Grant Nos. 10874232, 10774180, 90406010, and 60621091), by the
Chinese Academy of Sciences (Grant No.KJC-X2.YW.W09-5), and by
Chinese Department of Science and Technology (Grant No.
2005CB623602).
\end{acknowledgement}


\begin{thebibliography}{[1]}

\bibitem{Wolf}%
S.\,A. Wolf, D.\,D. Awschalom, R.\,A. Buhrman, J.\,M. Daughton,
S.~von Molnar, M.\,L. Roukes, A.\,Y. Chtchelkanova, and D.\,M.
Treger, Science \textbf{294}, 1488 (2001).

\bibitem{Pickett}%
W.\,E. Pickett and J.\,S. Moodera, Phys. Today \textbf{54}, 39
(2001).

\bibitem{Groot1}%
R.\,A. de Groot, F.\,M. Mueller, P.\,G. van Engen, and K.\,H. J.
Buschow, Phys. Rev. Lett. \textbf{50}, 2024 (1983).

\bibitem{Groot2}%
R.\,A. de Groot, Physica B \textbf{172}, 45 (1991).

\bibitem{Dong}%
J.\,W. Dong, L.\,C. Chen, C.\,J. Palmstrom, R.\,D. James, and
S.~McKernan, Appl. Phys. Lett. \textbf{75}, 1443 (1999).

\bibitem{Watts}%
S.\,M. Watts, S.~Wirth, S.~von Molnar, A.~Barry, and J.\,M. D.
Coey, Phys. Rev. B \textbf{61}, 9621 (2000).

\bibitem{Jedema1}%
F.\,J. Jedema, A.\,T. Filip, B.~van Wees, Nature \textbf{410}, 345
(2001).

\bibitem{Jedema2}%
S.~Soeya, J.~Hayakawa, H.~Takahashi, K.~Ito, C.~Yamamoto, A.~Kida,
H.~Asano, and M.~Matsui, Appl. Phys. Lett. \textbf{80}, 823
(2002).

\bibitem{Coey}%
J.\,M. D. Coey, M.~Viret, and S.~von Molnar, Adv. Phys.
\textbf{48}, 167 (1999).

\bibitem{sh}%
S.~Sanvito and N.\,A. Hill, Phys. Rev. B \textbf{62}, 15553
(2000).

\bibitem{Xie1}%
Y.\,-Q. Xu, B.\,-G. Liu, and D.\,G. Pettifor, Phys. Rev. B
\textbf{66}, 184435 (2002).

\bibitem{Xie2}%
B.\,-G. Liu, Phys. Rev. B \textbf{67}, 172411 (2003).

\bibitem{Xie3}%
W.\,-H. Xie, B.\,-G. Liu, and D.\,G. Pettifor, Phys. Rev. B
\textbf{68}, 134407 (2003).

\bibitem{lbg1}%
W.\,-H. Xie, Y.\,-Q. Xu, B.\,-G. Liu, and D.\,G. Pettifor, Phys.
Rev. Lett. \textbf{91}, 037204 (2003).

\bibitem{lbg2}%
B.\,-G. Liu, in: Half-metallic Alloys-Fundamentals and
Applications, edited by I Galanakis and P. H. Dederichs, Lecture
Notes in Physics Vol. \textbf{676}, (Springer, Berlin, 2005),
pp.267-291.

\bibitem{zbCT}%
M.\,G. Sreenivasan, K.\,L. Teo, M.\,B. A. Jalil, T.~Liew, T.\,C.
Chong, and A.\,Y. Du, IEEE Transactions on Magnetics \textbf{42},
2691 (2006).

\bibitem{Galanakis}%
I.~Galanakis, K.~Ozdogan, E.~Sasloglu, and B.~Aktas, Phys. Rev. B
\textbf{74}, 140408(R) (2006).

\bibitem{Fang}%
C.\,M. Fang, G.\,A. de Wijs, and R.\,A. de Groot, J. Appl. Phys.
\textbf{91}, 8340 (2002).

\bibitem{Durbin}%
S.\,M. Durbin, J.~Han, O.~Sungki, M.~Kobayashi, D.\,R. Menke, and
R.\,L. Gunshor, Appl. Phys. Lett. \textbf{55}, 2087 (1989).

\bibitem{mumnte}%
T.\,M. Giebultowicz, P.~Klosowski, N.~Samarth, H.~Luo, J.\,K.
Furdyna, and J.\,J. Rhyne, Phys. Rev. B \textbf{48}, 12817 (1993).

\bibitem{Li}%
Y.\,B. Li, Y.\,Q. Zhang, N.\,K. Sun, Q.~Zhang, D.~Li, J.~Li, and
Z.\,D. Zhang, Phys. Rev. B \textbf{72}, 193308 (2005).

\bibitem{Blaha}%
P.~Blaha, K.~Schwarz, P.~Sorantin, and S.\,B. Trickey, Comp. Phys.
Comm. \textbf{59}, 399 (1990).

\bibitem{Hohenberg}%
P.~Hohenberg and W.~Kohn, Phys. Rev. \textbf{136}, B864 (1964);
W.~Kohn and L.\,J. Sham, Phys. Rev. \textbf{140}, A1133 (1965).

\bibitem{Perdew}%
J.\,P. Perdew, K.~Burke, and M.~Ernzerhof, Phys. Rev. Lett.
\textbf{77}, 3865 (1996).

\bibitem{Hennion}%
B.~Hennion, W.~Szuszkiewicz, E.~Dynowska, E.~Janik, and
T.~Wojtowicz, Phys. Rev. B \textbf{66}, 224426 (2002).

\bibitem{Wei2}%
S.\,-H. Wei and A.~Zunger, Phys. Rev. B \textbf{48}, 6111 (1993).

\bibitem{Wei1}%
S.\,-H. Wei and A.~Zunger, Phys. Rev. Lett. \textbf{56}, 2391
(1986).

\bibitem{Nakamura}%
K.~Nakamura, T.~Ito, and A.\,J. Freeman, Phys. Rev. B \textbf{72},
064449 (2005).

\bibitem{Takahiro}%
S.~Takahiro, O.~Shintaro, K.~Takahiro, O.~Iwao, Nature
\textbf{437}, 1128 (2005).

\bibitem{Kubler}%
J.~Kubler, Phys. Rev. B \textbf{67}, 220403 (2003).

\end{thebibliography}
\end{document}